\documentclass[12pt,preprint]{aastex}
\begin{document}

\title{The RR Lyrae Distance to the Draco Dwarf Spheroidal Galaxy}

\author{A. Z. Bonanos, K. Z. Stanek, A.H. Szentgyorgyi, D.D. Sasselov,
G.\'{A}. Bakos}
\affil{Harvard-Smithsonian Center for Astrophysics, 60 Garden St.,
Cambridge, MA~02138}
\affil{\tt e-mail: abonanos@cfa.harvard.edu, kstanek@cfa.harvard.edu,
aszentgyorgyi@cfa.harvard.edu, sasselov@cfa.harvard.edu,
gbakos@cfa.harvard.edu}

\begin{abstract}

We present the first CCD variability study of the Draco dwarf
spheroidal galaxy. The data were obtained with the FLWO 1.2 m
telescope on 22 nights, over a period of 10 months, covering a
$22\arcmin\times22\arcmin$ field centered at
$\alpha=17\!\!:\!\!19\!\!:\!\!57.5, \delta=57\!\!:\!\!50\!\!:\!\!05,
{\rm J2000.0}$. The analysis of the $BVI$ images produced 163 variable
stars, 146 of which were RR Lyrae: 123 RRab, 16 RRc, 6 RRd and one
RR12. The other variables include a SX Phe star, four anomalous
Cepheids and a field eclipsing binary. Using the short distance scale
statistical parallax calibration of Gould \& Popowski and 94 RRab
stars from our field, we obtain a distance modulus of $\rm
(m-M)_{0}=19.40 \pm 0.02\, (stat) \pm 0.15 \,(syst)$ mag for Draco,
corresponding to a distance of 75.8 $ \rm \pm 0.7 \,(stat) \pm 5.4
\,(syst)$ kpc. By comparing the spread in magnitudes of RRab stars in
$B,V$ and $I$, we find no evidence for internal dust in the Draco
dwarf spheroidal galaxy.

The catalog of all variables, as well as their photometry and finding
charts, is available electronically via {\tt anonymous ftp} and the
{\tt World Wide Web}. The complete set of the CCD frames is available
upon request.

\end{abstract}

\section{Introduction}

Dwarf spheroidal (dSph) galaxies are probably the most common types of
galaxies in the present-day Universe. They are metal poor galaxies
with a metallicity $Z < 0.001$ \citep{Mat98}, which resembles that
found in galactic globular clusters. Most dSph galaxies show evidence
for multiple star-formation episodes, having populations of different
ages. There are very few, namely Tucana, Draco and possibly Ursa
Minor, that host a single stellar population older than 10 Gyr
\citep[see][and references therein]{Dal03}.

The Draco dSph galaxy, a companion to the Milky Way, was discovered by
\citet{Wil55} and was first observed by \citet{Baa61} for
variables. They found 261 variables in their $24\arcmin \times
24\arcmin$ field, but only measured 137 for magnitudes. Of these, 133
were RR Lyrae variables, which they used to derive the distance to the
galaxy. There have not been any recent variability studies of Draco,
except for the survey by \citet{Kin02} which is currently
underway. The lack of high quality CCD observations of the RR Lyrae in
Draco dSph motivated us to do this project.

However, several studies of Draco's stellar population have been
conducted and for these CCD photometry has been obtained.
\citet{Gri98} present the CMD diagram obtained from observations with
the {\em Hubble Space Telescope (HST)}\/ and confirm that star
formation in Draco was primarily single-epoch and that Draco is very
similar to the globular clusters M68 and M92, but 1.6 Gyr older. It
has a luminosity of $2 \times 10^{5}\, \rm L_{\sun}$, which places it
among the least luminous galaxies known. \citet{Bel02} have done a
comparative study of the Draco and Ursa Minor dSph galaxies with new
$V,I$ photometry.  Recently, \citet{Rav03} have released a catalog of
photometry of $\sim 5,600$ stars in Draco. They find 142 candidate
variables from their colors, using photometry from five
catalogs. However, a uniform dataset taken with the same instrument
would be more reliable for finding RR Lyrae and obtaining accurate
photometry and periods for them.

In this paper, we present a catalog of variable stars found in Draco
dSph. The paper is organized as follows: \S 2 provides a description
of the observations; the data reduction procedure, calibration and
astrometry is outlined in \S 3; the catalog of variable stars is
presented in \S 4. In \S 5 we determine the distance to Draco and in
\S 6 we summarize our results.

\section{Observations}
The observations of the Draco dSph were made with the 1.2m telescope
at the Fred Lawrence Whipple Observatory on Mount Hopkins, Arizona,
between August 19th, 1998 and June 20th, 1999, over 22 nights. We used
the ``4Shooter'' camera \citep{Sze03}, with 4~thinned and AR-coated
Loral 2048$^2$ pixel CCDs. The pixels are 15 microns in size and map
to $0.33\arcsec$ per pixel on the focal plane, making each image
$11\arcmin$ on the side. The camera was centered at
$\alpha=17\!\!:\!\!19\!\!:\!\!57.5, \delta=57\!\!:\!\!50\!\!:\!\!05,
{\rm J2000.0}$. The data consists of 148 $\times$ 600 $s$ exposures in
the $V$ filter, 44 $\times$ 900 $s$ exposures in the $B$ filter and 47
$\times$ 600 $s$ exposures in the $I$ filter. The median value of the
seeing in $V$ was $2.0\arcsec$. The field was observed through
airmasses ranging from 1.12 to 1.55, with the median being 1.19. The
completeness of our photometry starts to drop rapidly at about 22.5 in
$I$ and 23 mag in $V$ and $B$. The CCDs saturate for stars brighter
than 14 in $I$, 15 in $V$ and 15.5 mag in $B$. On one photometric
night of the run, several images of standard \citet{Lan92} fields were
taken.

\section{Data Reduction, Calibration and Astrometry}

Preliminary processing of the data was performed with standard
routines in the IRAF~\footnote{IRAF is distributed by the National
Optical Astronomy Observatories, which are operated by the Association
of Universities for Research in Astronomy, Inc., under cooperative
agreement with the NSF.} CCDPROC package. The differential photometry
for the variable stars was extracted using the ISIS image subtraction
package \citep{Ala98,Ala00} from the $V$-band data. The
DAOPHOT/ALLSTAR package \citep{Ste87} was used for the conversion into
magnitudes. \citet{Moc01} describe the procedure in detail.

On August 31st, 1998, we observed 2 sets of 3 \citet{Lan92} fields in
the $BVI$ filters at air masses ranging from 1.18 to 1.99. The
transformation from the instrumental to the standard system was
derived for each chip in the following form:

\begin{eqnarray*}
b = B + \rm \chi_{b} + \rm \xi_{b} \cdot(B-V) + \rm \kappa_{b} \cdot X\\
v = V + \rm \chi_{v1} + \rm \xi_{v1} \cdot(B-V) + \rm \kappa_{v1} \cdot X\\
v = V + \rm \chi_{v2} + \rm \xi_{v2} \cdot(V-I) + \rm \kappa_{v2} \cdot X\\
i = I + \rm \chi_{i} + \rm \xi_{i} \cdot(V-I) + \rm \kappa_{i} \cdot X\\
\end{eqnarray*}

\noindent where lowercase letters correspond to the instrumental
magnitudes, uppercase letters to standard magnitudes, X is the
airmass, $\chi$ is the zeropoint, $\xi$ the color and $\kappa$ the
airmass coefficient. Since most of the color coefficients are small,
we used $B-V=V-I=1$ when transforming the magnitudes of our
stars. Note that the $B$-band coefficients are larger, therefore our B
magnitudes for red stars may be off by 0.1 mag or 0.2 mag (in chip 2),
in the worst case.

\citet{Ste00} has calibrated $\sim400$ stars in the Draco dSph as
secondary standards. In chips 3 and 4, where our overlap was large, we
normalized to his photometry using the brightest 80 stars (to 19.5
mag) and 66 stars (to 20th mag) respectively to determine the offsets
in $V$. The difference between his photometry and ours in these chips
was 0.04 and 0.02 mag. In chips 1 and 2 the overlap was too small for
a meaningful comparison, thus we kept our own photometry. We then
compared our normalized $V$ photometry in chips 3 and 4 with the
photometry of \citet{Gri98} from the {\em HST}\/. For 100 stars down
to 19 mag, the differences between their photometry and ours were 0.06
and 0.03 mag. \citet{Bel02} have obtained ($V,I$) photometry of the
field and the agreement with our photometry is good, the largest
offset being 0.07 mag in chip 3, $V$-band.

Equatorial coordinates were determined for the $V$ star list. The
transformation from rectangular to equatorial coordinates was derived
using for chips 1-4: 174, 146, 400 and 282 transformation stars
respectively with $V<20$ from the USNO-A2.0 \citep{Mon96} catalog. The
median difference between the catalog and the computed coordinates for
the transformation stars was $0.\arcsec3$ in RA and $0.\arcsec3$ in
Dec. We also compared the astrometry to Stetson's catalog and found
18, 4, 272 and 151 matches for chips 1-4, having a median offset
$<0.\arcsec2$. We use these derived J2000.0 equatorial coordinates to
name the variables in the format: Draco$hhmmss.s$+$ddmmss.s$. The
first three fields ($hhmmss.s$) correspond to RA expressed in hours,
the last three ($ddmmss.s$) to Dec, expressed in degrees, separated by
the declination sign.

\section{Results}

\begin{figure}
\plotone{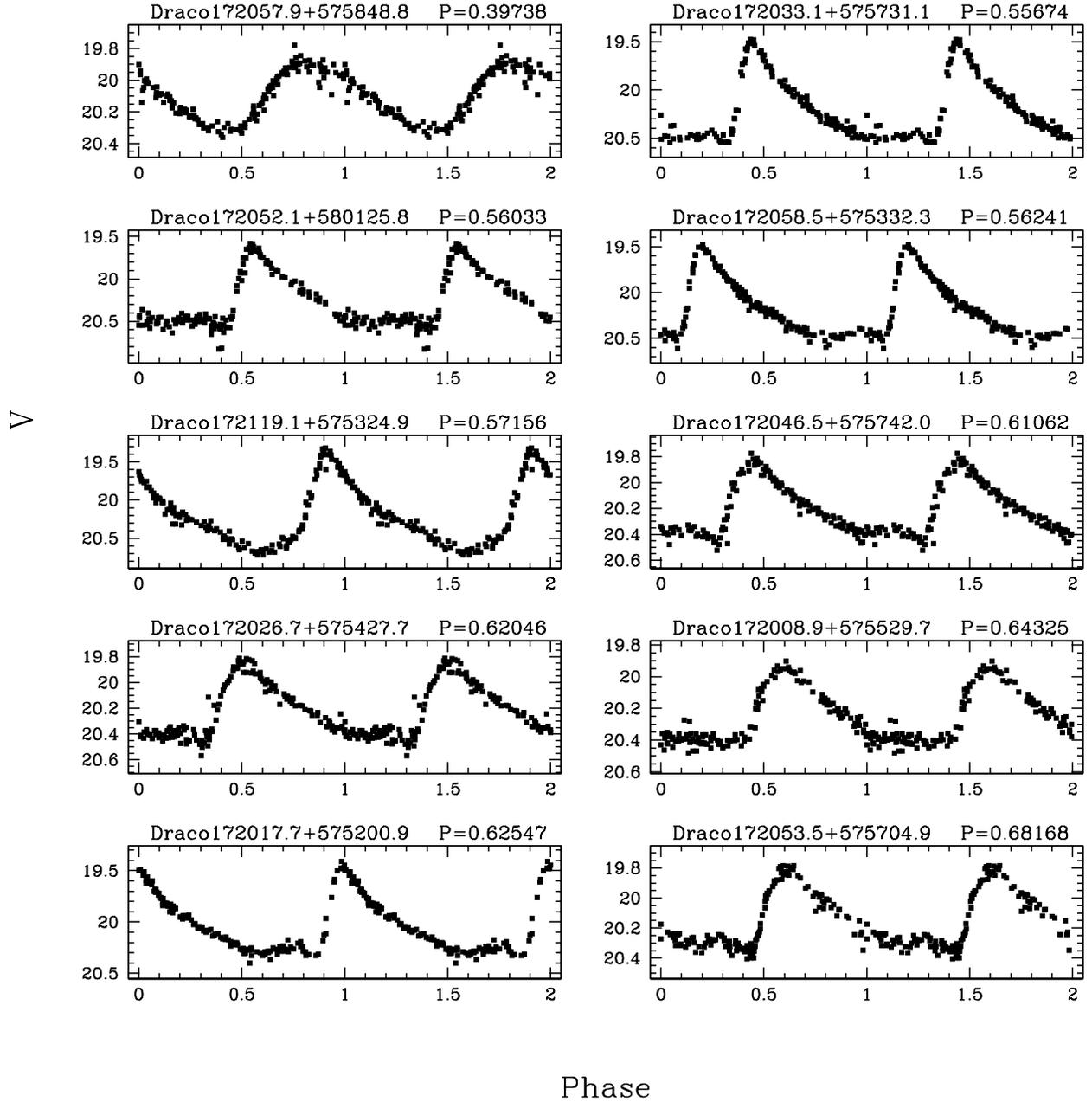}
\caption{Sample light curves of RR Lyrae variables found in Draco
dSph, representing typical quality data over a range of periods.}
\label{rrlyrlc}
\end{figure}

\begin{figure}
\plotone{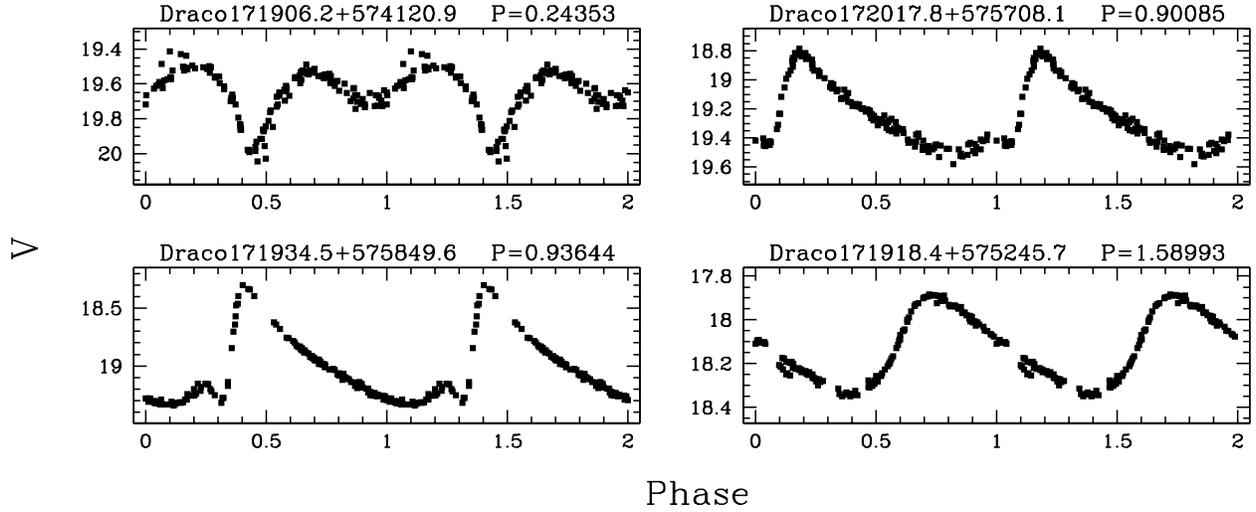}
\caption{Light curves of selected other periodic variables found in
Draco dSph. An eclipsing binary and 3 anomalous Cepheids are shown.}
\label{perlc}
\end{figure}

\begin{figure}
\plotone{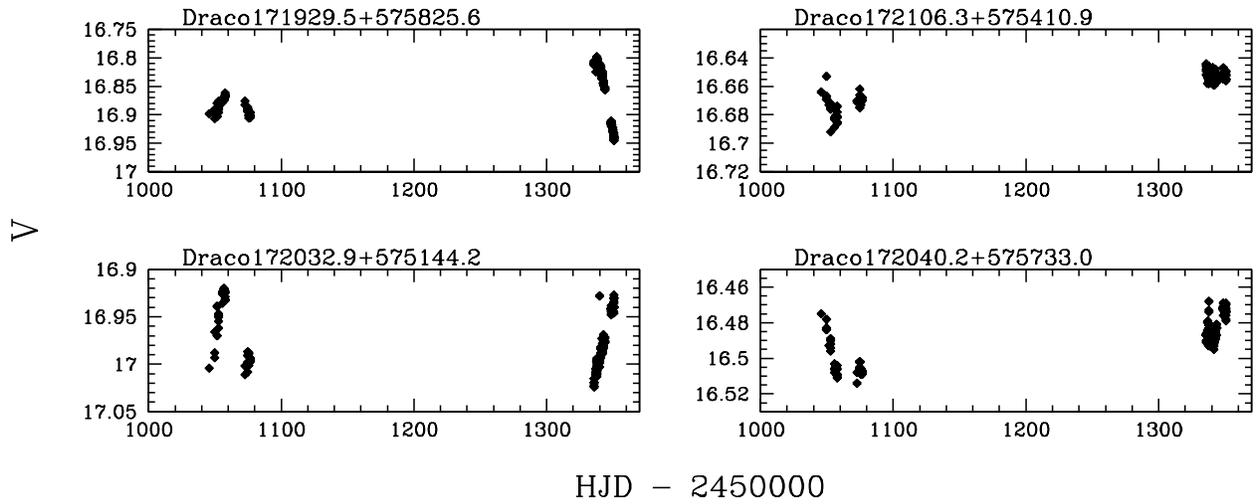}
\caption{Sample light curves of other variables found in Draco dSph.}
\label{operlc}
\end{figure}

Our search for variables in our field in Draco produced 163 stars, 136
of which were previously identified by \citet{Baa61}. The remaining 27
are new discoveries. Of these 163 stars, 146 are RR Lyrae, 4 are
anomalous Cepheids and the remaining 13 are other long period or
non-periodic variables. We found a new field eclipsing binary and a SX
Phe star among these. Of the 146 RR Lyrae, 123 are fundamental mode
pulsators (RRab), 16 are first overtone pulsators (RRc), 6 are
double-mode pulsators (RRd) and one is pulsating in the first and
second overtone (RR12). Figures~\ref{rrlyrlc}, \ref{perlc}
and~\ref{operlc} show typical light curves of RR Lyrae and other
variables found in Draco. Tables~\ref{rrlyr} and~\ref{var} present
coordinates, periods, intensity averaged $BVI$ magnitudes, $V$-band
amplitudes, the type of variable and the corresponding name given in
\citet{Baa61}. Stars exhibiting the Blazhko effect are also
marked. The catalog of all variables, as well as their $BVI$
photometry and $V$ finding charts, is available electronically via
{\tt anonymous ftp}\footnote{On {\tt cfa-ftp.harvard.edu}, in {\tt
pub/kstanek/DIRECT/Draco}} and the {\tt World Wide Web}\footnote{{\tt
http://cfa-www.harvard.edu/\~\/kstanek/DIRECT/Draco}}. The complete
set of the CCD frames is available upon request.

We used the multiharmonic analysis of variance technique \citep{Sch96}
to search the light curves for periodicity. Additionally, Fourier
series were fit to the RR Lyrae light curves phased to the period
determined earlier and parameters such as the amplitude of the
variation and amplitude and phase of each harmonic were calculated. We
searched for multiperiod variables by subtracting the first three
harmonics of the Fourier series from the phased light curves and then
repeating the period search. We then redetermined secondary periods
for the 6 double mode (RRd) stars found by \citet{Nem85} in the data
of \citet{Baa61}, by searching the periodogram where the second period
was expected.

\citet{Baa61} do not find any red irregular or long period variables
in their data, with the exception of BS-203, a bright blue variable
with a period of $\sim 3$ years. We observed all of their ``special
variables'' except for BS-138. The only significant difference in
these is that the period we find for BS-134 is 0.592 days, not 1.458
days, which agrees with \citet{Nem85} who realanyzed the data of
\citet{Baa61}. The periods we calculated for variables also found by
\citet{Baa61} are very similar in most cases.  The cases that differ
are marked with an asterisk in Table~\ref{rrlyr}. As a result some
stars are classified differently from \citet{Baa61}. BS-97, BS-121,
BS-173 and BS-145 are all RRc stars and BS-190, BS-169, BS-143, BS-72,
BS-11, BS-112 are RRd stars. We did not find variables BS-10, BS-31,
BS-111 and BS-195 due to the proximity of highly saturated stars.

\begin{figure}
\plotone{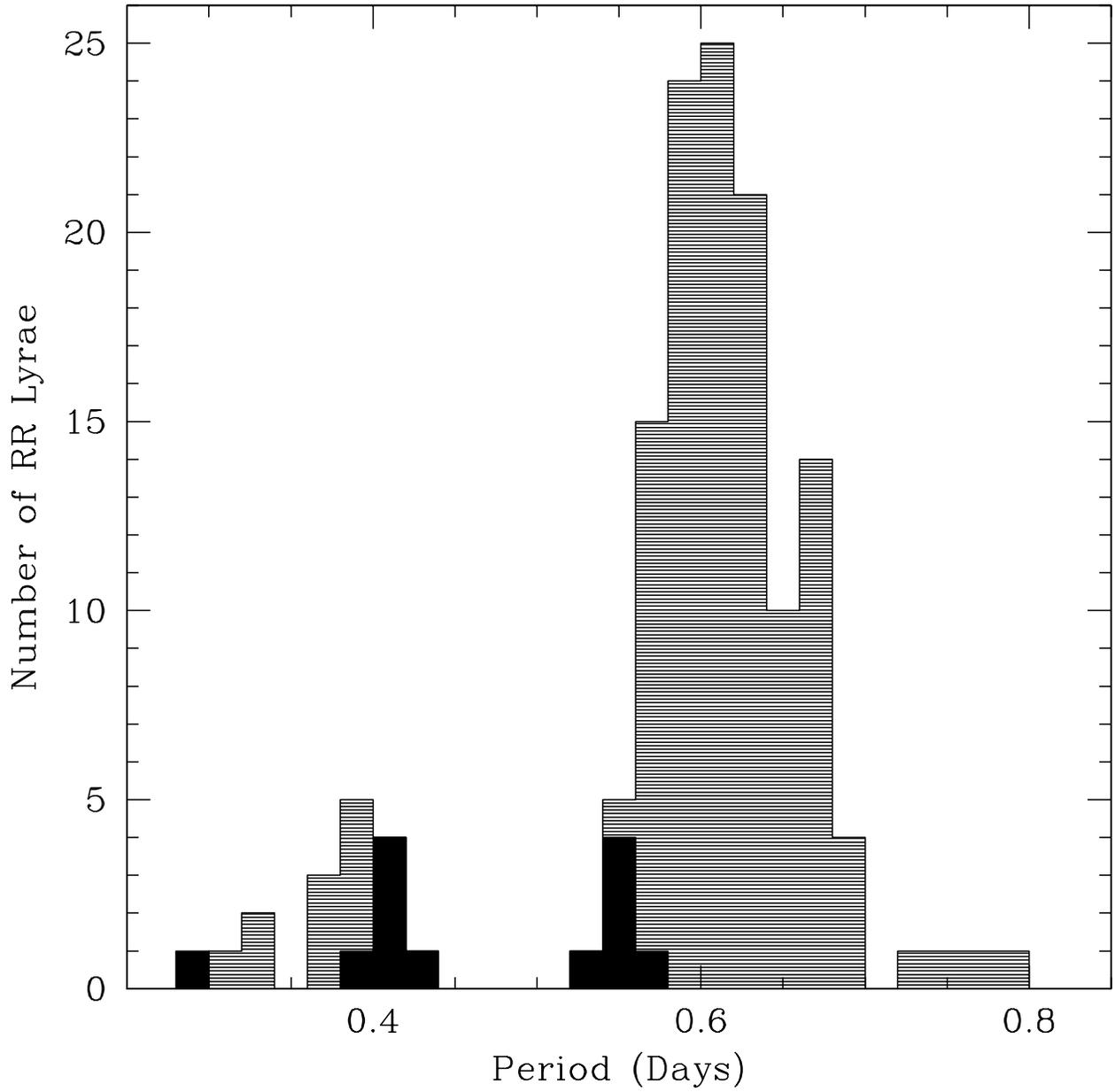}
\caption{Period distribution of 146 RR Lyrae in Draco. The median
period for RRab stars is 0.617 and for RRc stars 0.392. Both
components of the double-mode stars are also plotted in black.}
\label{histogram}
\end{figure}

\begin{figure}
\plotone{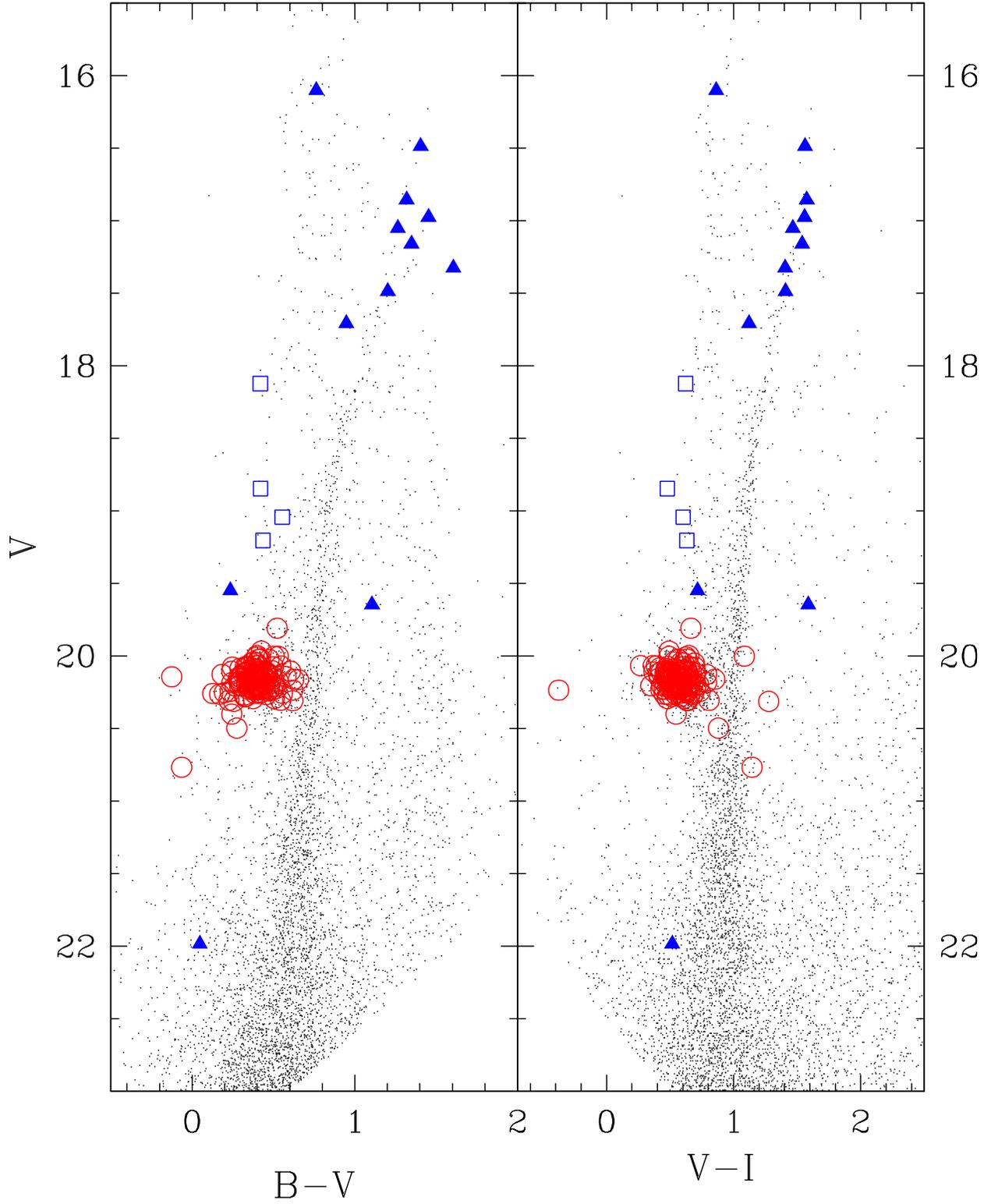}
\caption{CMD for variables and nonvariable stars. Circles represent RR
Lyrae, squares are anomalous Cepheids and triangles are other
variables.}
\label{cmd}
\end{figure}

\begin{figure}
\plotone{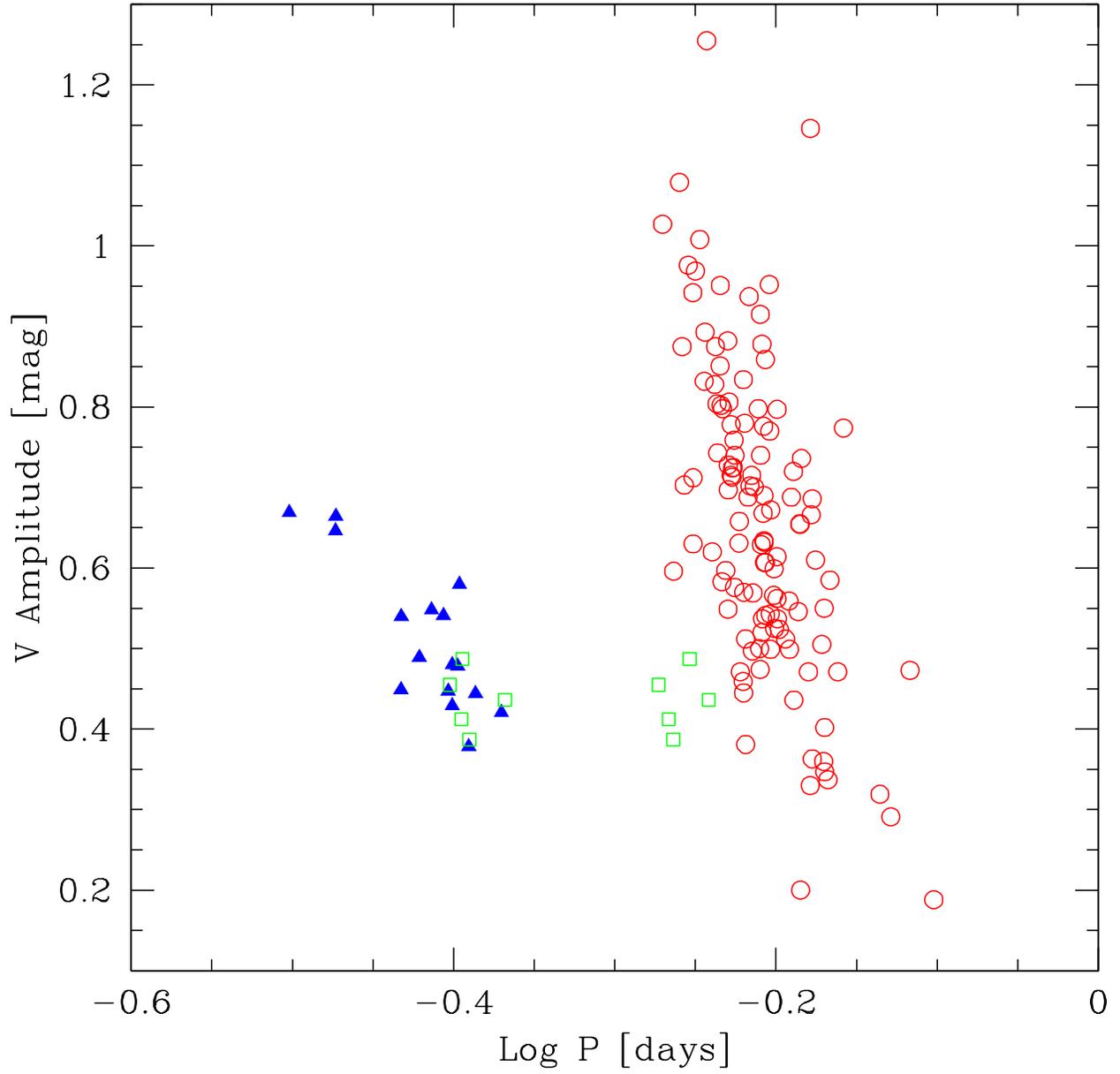}
\caption{Period-amplitude relation for 146 RR Lyrae in Draco
dSph. Circles represent RRab stars, triangles RRc stars and squares
RRd stars, for which both periods are plotted.}
\label{peramp}
\end{figure}

We present a histogram of the 139 RRab and RRc Lyrae in Draco, with
0.02 day bins in Figure~\ref{histogram}. Both components of the
double-mode stars are also plotted (in black). The median period for
RRab stars is 0.617 days and for RRc stars is 0.392 days, which places
the Draco dSph between Oosterhoff type I ($\sim 0.55$ days) and type
II ($\sim 0.65$ days) clusters, similarly to other dSph \citep{Dal03}.
In Figure~\ref{cmd} we present a color magnitude diagram (CMD) of
stars in Draco. Circles represent RR Lyrae, squares are anomalous
Cepheids and triangles are other variables. Among these other
variables is the long period blue variable BS-203, a foreground 0.24
day eclipsing binary, and a multimode SX Phe star which is pulsating
in three modes, with periods 0.068, 0.073 and 0.079 days. The
period-amplitude diagram for the 146 RR Lyrae in Draco is shown in
Figure~\ref{peramp}.  Circles represent RRab stars, triangles RRc
stars and squares RRd stars, for which both periods are plotted.

\section{Distance to Draco dSph}

The distance to the Draco dSph galaxy has been estimated by several
authors. \citet{Baa61} obtained a distance of $d=99$ kpc assuming an
absolute magnitude of $M_{B}=0.5$ mag for RR Lyrae; \citet{Nem85}
obtained $d=84 \pm12$ kpc using RRd stars found by reanalyzing the
data of \citet{Baa61}; \citet{Apa01} obtained $d=80 \pm7$ kpc by the
magnitude of the horizontal branch at the RR Lyrae instability strip;
\citet{Bel02} obtained $d=92.9 \pm 6$ kpc by fitting template cluster
horizontal branches.

We use the short distance scale statistical parallax calibration of
\citet{Gou98}, which is a robust method of measuring the absolute
magnitude of RR Lyrae stars. They find

\begin{equation}
\rm M_{V} = 0.77 \pm 0.13, 
\end{equation}

\noindent at $\langle [\rm Fe/H] \rangle=-1.60$, for a sample of 147
halo RR Lyrae stars with high-quality proper motions from the
\citet{Hip97} and \citet{Kle93} surveys. \citet{She01} find $\langle
[\rm Fe/H] \rangle=-2.00 \pm 0.21$ in Draco dSph from high resolution
spectroscopy of 6 red giants in the galaxy. \citet{Leh92} also found a
metallicity of $\langle [\rm Fe/H] \rangle=-1.9$, $\sigma=0.4$, from
spectra of 14 giants. The abundances seem to fall into two groups, one
with an average [Ca, Mg/H] near $-1.6 \pm 0.2$ and the other $-2.3 \pm
0.2$. We adopt the value $\langle [\rm Fe/H] \rangle=-2.00$
\citep[also quoted in][]{Mat98} for our distance determination. The
metallicity correction is derived from the slope of the
luminosity-metallicity relation for RR Lyrae, which lies between 0.2
\citep{Cha99} and 0.3 \citep{San93}. Using 0.2 for the value of the
slope and 0.4 dex for the difference in metallicity of Draco dSph from
galactic RR Lyrae, we find $M_{V}=0.69$ for the RR Lyrae in Draco
dSph.

Draco dSph is located at Galactic coordinates $l=86\arcdeg\!\!.37,
b=34\arcdeg\!\!.72$. To remove the effects of the Galactic
interstellar extinction we used the reddening map of \citet{Sch98}
which yields $E(B-V)=0.027$ mag. This corresponds to expected values
of Galactic extinction of $A_I=0.053$, $A_V=0.091$, $A_B=0.118$ mag,
using the extinction corrections of \citet{Car89} as prescribed in
\citet{Sch98}. 

\begin{figure}
\plotone{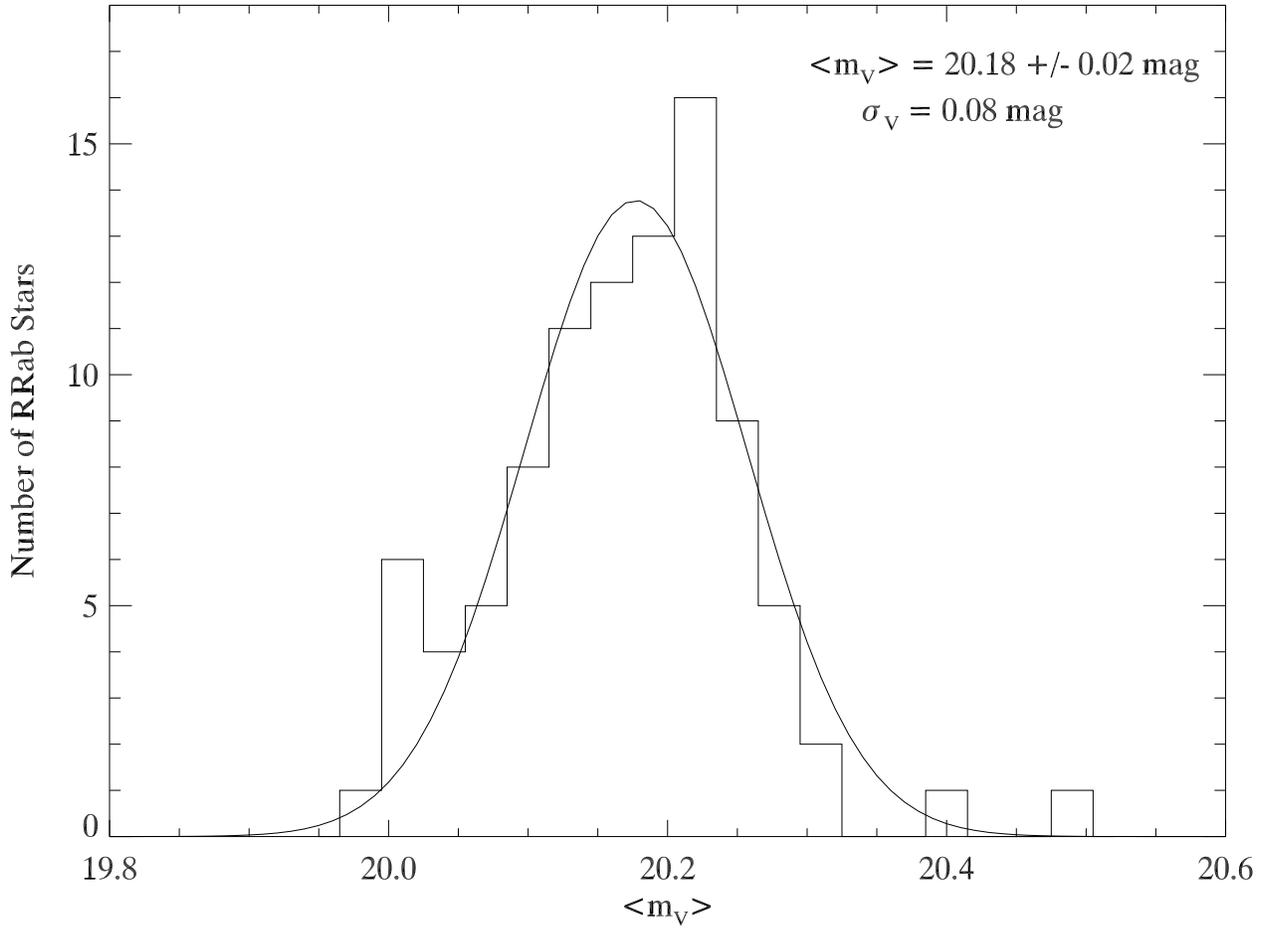}
\caption{Histogram of 94 RRab magnitudes and the Gaussian fit,
centered at $\rm \langle m_{V} \rangle=20.18 \pm 0.02$ mag.}
\label{gauss}
\end{figure}

For the distance determination we only used the RRab stars found in
chips 3 and 4, which are normalized to the photometry of
\citet{Ste00}. There are 94 such stars in our data. The remaining RRab
stars from chip 1 and 2 are not included in this list. We fit a
Gaussian to a histogram of these 94 stars, using 0.03 mag bins and
found $\rm \langle m_{V} \rangle = 20.18 \pm 0.02$, $\sigma= 0.08$, as
shown in Figure~\ref{gauss}. This value is in agreement with the value
of \citet{Apa01} for the horizontal branch at the RR Lyrae instability
strip, $20.2 \pm 0.1$ mag, and with the value of \citet{Bel02} of
$20.30 \pm 0.12$ mag, obtained by fitting to the template cluster
M68. Our measurement implies a distance modulus of $\rm
m_{V}-M_{V}=19.49$ mag. Correcting for extinction gives a true
distance modulus of $\rm (m-M)_{0}=19.40$ mag and a distance of 75.8
kpc to Draco dSph.

The systematic errors are 0.06 mag in $A_V$, 0.03 mag in photometry,
0.13 mag in the calibration method and 0.04 mag in metallicity. The
error in the reddening from \citet{Sch98} is 0.02 mag, which
corresponds to 0.06 mag in $A_V$, and the error in metallicity is
calculated from a conservative error of 0.1 in the slope of the
luminosity-metallicity relation times 0.4 dex, the metallicity
difference. Adding the systematic errors in quadrature gives a
conservative total estimate of 0.15 mag, which is dominated by the
calibration error. We consider the effects of internal extinction to
be negligible from a comparison of the spread in magnitudes of RRab
stars in different filters. Similarly to Figure~\ref{gauss} for $V$
with $\sigma= 0.08$, the spread in magnitudes of RRab stars in $B$ and
$I$ are $\sigma= 0.10$ and $0.12$, respectively. Thus we find no
evidence for internal dust in the Draco dSph galaxy.

The statistical error is 0.02 mag, which leads to a true
distance modulus of $\rm (m-M)_{0}=19.40 \pm 0.02\, (stat)\pm
0.15\,(syst)$ mag, corresponding to a distance of $\rm 75.8 \pm
0.7\,(stat)\pm 5.4\,(syst)$ kpc.

\section{Conclusions}
We have presented the results of the first CCD variability study in
the Draco dSph galaxy since \citet{Baa61}. Our search produced 163
variable stars, 146 of which are RR Lyrae, 4 are anomalous Cepheids, 1
is a field eclipsing binary, 1 a SX Phe star and 11 are other types of
variables. We have used the short distance scale statistical parallax
calibration of \citet{Gou98} for 94 RRab in our field and obtained a
distance modulus of $\rm (m-M)_{0}=19.40 \pm 0.02\, (stat)\pm
0.15\,(syst)$ mag. By comparing the spread in magnitudes of RRab stars
in different filters, we find no evidence for internal dust in the
Draco dSph galaxy.

The catalog of all variables, as well as their photometry and finding
charts, is available electronically via {\tt anonymous ftp} and the
{\tt World Wide Web}. The complete set of the CCD frames is available
upon request.

\acknowledgments{We thank Grzegorz Pojma\'nski and Wojtek Pych for
their LC and Fourier series programs and Barbara Mochejska for her
help. We also thank Janusz Kaluzny, Piotr Popowski, John Huchra and
the referee for their careful reading of and comments on the
manuscript. We finally thank Mike Pahre, Emilio Falco, Lucas Macri and
Saurabh Jha, for helping obtain the observations. Guest User, Canadian
Astronomy Data Centre, which is operated by the Dominion Astrophysical
Observatory for the National Research Council of Canada's Herzberg
Institute of Astrophysics.}

\begin{deluxetable}{clcccccl}
\tabletypesize{\footnotesize}
\tablewidth{0pc}
\tablecaption{\sc RR Lyrae in Draco Dwarf Galaxy}
\tablehead{
\colhead{} & \colhead{$P$} & \colhead{$\langle V \rangle$} &\colhead{$\langle I \rangle$} & \colhead{$\langle B \rangle$} & \colhead{$\rm Amp_{V}$} & \colhead{} & \colhead{}\\
\colhead{Name} & \colhead{($days$)} & \colhead{(mag)} & \colhead{(mag)} & \colhead{(mag)} & 
\colhead{(mag)} & \colhead{Type} & \colhead{Comments\tablenotemark{a}}}
\startdata
Draco172052.0+575532.4 &  0.28767$^{\dag}$ & 20.205& 19.855& 20.496 & 0.356 & RR12     & BS-170\\
 	\nodata	       &  0.40431 	   &\nodata&\nodata&\nodata &\nodata&\nodata& \nodata\\
Draco172059.1+580006.0 &  0.31478* 	   & 20.263& 19.808& 20.432 & 0.669 & c     & BS-97\\
Draco171936.4+574930.0 &  0.33633 	   & 20.230& 19.802& 20.627 & 0.646 & c     & BS-46\\
Draco171936.1+575415.8 &  0.33646* 	   & 20.081& 19.659& 20.417 & 0.664 & c     & BS-121\\
Draco171908.1+575835.0 &  0.36924$^{\dag}$ & 20.204& 19.769& 20.576 & 0.449 & c     & BS-110\\
Draco172059.5+575542.7 &  0.36947* 	   & 20.151& 19.735& 20.517 & 0.540 & c     & BS-173\\
Draco171938.4+574724.7 &  0.37903$^{\dag}$ & 20.152& 19.683& 20.570 & 0.489 & c     & BS-50 \\
Draco172112.7+580131.2 &  0.38572$^{\dag}$ & 20.176& 19.726& 20.462 & 0.548 & c     & BS-181\\
Draco172110.8+574736.2 &  0.39233$^{\dag}$ & 20.068& 19.675& 20.613 & 0.541 & c     & BS-179\\
Draco171917.5+580107.5 &  0.39502 	   & 20.145& 19.671& 20.433 & 0.447 & c     & \nodata\\
Draco172042.5+575153.5 &  0.39597* 	   & 20.120& 19.617& 20.509 & 0.455 & d     & BS-190\\
 	\nodata	       &  0.53351 	   &\nodata&\nodata&\nodata &\nodata&\nodata& \nodata\\
Draco171917.5+574843.0 &  0.39728$^{\dag}$ & 20.120& 19.640& 20.580 & 0.480 & c     & BS-191\\
Draco172057.9+575848.8 &  0.39738* 	   & 20.071& 19.589& 20.408 & 0.429 & c     & BS-145\\
Draco171942.4+575837.9 &  0.40049$^{\dag}$ & 20.108& 19.607& 20.448 & 0.478 & c     & BS-120\\
Draco172017.5+574601.7 &  0.40155$^{\dag}$ & 20.151& 19.624& 20.604 & 0.580 & c     & BS-153\\
Draco172042.3+575852.7 &  0.40258* 	   & 20.126& 19.591& 20.310 & 0.412 & d     & BS-169\\
 	\nodata	       &  0.54144 	   &\nodata&\nodata&\nodata &\nodata&\nodata& \nodata\\
Draco171942.5+575449.8 &  0.40319* 	   & 20.129& 19.594& 20.519 & 0.487 & d     & BS-143\\
 	\nodata	       &  0.55789	   &\nodata&\nodata&\nodata &\nodata&\nodata& \nodata\\
Draco172119.5+575236.0 &  0.40685$^{\dag}$ & 20.116& 19.632& 20.448 & 0.378 & c     & BS-131\\
Draco172012.4+575412.0 &  0.40720* 	   & 20.105& 19.614& 20.710 & 0.387 & d     & BS-72\\
  	\nodata	       &  0.54489 	   &\nodata&\nodata&\nodata &\nodata&\nodata& \nodata\\
Draco171930.5+575633.7 &  0.41080$^{\dag}$ & 20.110& 19.681& 20.512 & 0.444 & c     & BS-182\\
Draco172041.9+575827.6 &  0.41215* 	   &\nodata& 19.616& 20.310 &\nodata& d     & BS-11\\
  	\nodata	       &  0.55075 	   &\nodata&\nodata&\nodata &\nodata&\nodata& \nodata\\
Draco171907.8+574432.7 &  0.42626 	   & 20.122& 19.635& 20.510 & 0.421 & c     & \nodata\\
Draco172106.4+575153.4 &  0.42842* 	   & 20.115& 19.482& 20.442 & 0.436 & d     & BS-112\\
  	\nodata	       &  0.57322 	   &\nodata&\nodata&\nodata &\nodata&\nodata& \nodata\\
Draco172047.2+575759.7 &  0.53656 	   & 20.110 & 19.736 &  20.449  & 1.027 & ab    & BS-13\\
Draco172017.0+575240.2 &  0.54513	   & 20.497 & 19.617 &  20.772  & 0.596 &ab,Bl? & BS-34\\
Draco171919.5+575738.9 &  0.54967	   & 20.280 & 19.700 &  20.604  & 1.079 & ab    & BS-18\\
Draco171923.3+575555.9 &  0.55185	   & 20.002 & 18.917 &  20.506  & 0.875 & ab    & \nodata\\
Draco172008.4+575203.7 &  0.55345	   & 20.181 & 19.653 &  20.474  & 0.703 & ab,Bl & BS-37\\
Draco172033.1+575731.1 &  0.55674	   & 20.156 & 19.664 &  20.441  & 0.976 & ab    & BS-124\\
Draco172052.1+580125.8 &  0.56033	   & 20.212 & 19.698 &  20.573  & 0.942 & ab    & BS-94\\
Draco172058.9+575344.6 &  0.56055	   & 20.266 & 19.783 &  20.720  & 0.630 & ab    & BS-163\\
Draco172032.5+575509.8 &  0.56062	   & 20.244 & 19.508 &  20.477  & 0.712 & ab    & BS-21\\
Draco172049.8+575405.5 &  0.56174	   &\nodata & 19.530 &  20.615  &\nodata& ab    & BS-25\\
Draco172058.5+575332.3 &  0.56241	   & 20.119 & 19.660 &  20.546  & 0.969 & ab    & BS-175\\
Draco172042.5+573955.7 &  0.56600	   & 20.142 & 19.597 &  20.719  & 1.008 & ab    & \nodata\\
Draco172016.7+575312.0 &  0.56917	   &\nodata &\nodata & \nodata  &\nodata& ab    & BS-29\\
Draco172015.2+575917.5 &  0.56957	   & 20.195 & 19.663 &  20.532  & 0.832 & ab    & BS-8\\
Draco172020.5+580056.3 &  0.57011	   & 20.252 & 19.625 &  20.451  & 0.893 & ab    & BS-5\\
Draco172119.1+575324.9 &  0.57156$^{\dag}$ & 20.067 & 19.799 &  20.459  & 1.255 & ab    & BS-130\\
Draco172038.4+575236.1 &  0.57603	   & 20.270 & 19.593 &  20.592  & 0.620 & ab,Bl & BS-35\\
Draco171927.1+574653.5 &  0.57631$^{\dag}$ & \nodata& 19.693 & \nodata 	&\nodata& ab    & BS-116\\
Draco172041.9+575750.5 &  0.57642	   & \nodata& 19.732 &  20.572  &\nodata& ab    & BS-12\\
Draco171926.6+575334.2 &  0.57804	   & 20.279 & 19.764 &  20.610  & 0.828 & ab    & BS-15\\
Draco172042.9+575129.2 &  0.57877	   & 20.219 & 19.683 &  20.573  & 0.875 & ab,Bl & BS-41\\
Draco171934.1+575535.8 &  0.58002	   & 20.107 & 19.562 &  20.508  & 0.804 & ab    & BS-107\\
Draco171941.4+575327.6 &  0.58048$^{\dag}$ & 20.249 & 19.691 & \nodata  & 0.743 & ab    & BS-22\\
Draco171939.1+575803.9 &  0.58258$^{\dag}$ & 20.160 & 19.568 &  20.606  & 0.851 & ab    & BS-102\\
Draco171949.9+574904.5 &  0.58273	   & 20.101 & 19.633 &  20.592  & 0.951 & ab    & BS-48\\ 
Draco171935.7+575832.2 &  0.58332$^{\dag}$ & 20.206 & 19.571 &  20.623  & 0.802 & ab    & BS-76\\
Draco172103.5+575950.9 &  0.58410	   & 20.124 & 19.653 &  20.508  & 0.583 & ab,Bl & BS-96\\
Draco171948.2+575451.6 &  0.58468$^{\dag}$ & 20.253 & 19.622 &  20.674  & 0.798 & ab    & BS-73\\
Draco171942.9+575527.1 &  0.58722	   & 20.201 & 19.495 &  20.612  & 0.597 & ab,Bl & BS-147\\
Draco171931.8+575705.0 &  0.58886	   & 20.259 & 19.680 &  20.646  & 0.882 & ab    & BS-144\\
Draco172011.5+575802.9 &  0.58921	   & 20.106 & 19.490 &  20.350  & 0.549 & ab,Bl & BS-123\\
Draco172059.3+580126.6 &  0.58938$^{\dag}$ & 20.065 & 19.697 &  20.473  & 0.697 & ab    & BS-118\\
Draco171939.9+575753.5 &  0.58939	   & 20.209 & 19.636 &  20.609  & 0.728 & ab,Bl & BS-196\\
Draco171924.8+575847.2 &  0.59003	   & 20.219 & 19.628 &  20.652  & 0.806 & ab,Bl & BS-129\\
Draco171858.2+575256.6 &  0.59182	   & 20.041 & 19.510 &  20.419  & 0.778 & ab    & BS-104\\
Draco171953.4+574844.9 &  0.59191	   & 20.007 & 19.273 &  20.459  & 0.715 & ab    & BS-84\\ 
Draco172047.3+575523.5 &  0.59256	   & 20.235 & 19.714 &  20.615  & 0.713 & ab    & BS-185\\
Draco172008.9+575623.3 &  0.59285	   & 20.166 & 19.671 &  20.510  & 0.724 & ab    & BS-126\\
Draco171902.2+574754.6 &  0.59369$^{\dag}$ & 20.241 & 19.605 &  20.708  & 0.725 & ab    & BS-115\\
Draco172113.0+575351.1 &  0.59441	   & 20.136 & 19.630 &  20.512  & 0.759 & ab    & BS-189\\
Draco171912.0+575437.2 &  0.59466$^{\dag}$ & 20.151 & 19.588 &  20.785  & 0.576 & ab    & BS-109\\
Draco172107.3+575800.8 &  0.59507	   & 20.174 & 19.615 &  20.525  & 0.740 & ab    & BS-183\\
Draco172040.3+575604.5 &  0.59854	   & 20.220 & 19.723 &  20.622  & 0.631 & ab    & BS-17\\
Draco171849.6+575356.1 &  0.59875	   & 20.147 & 19.566 &  20.536  & 0.658 & ab    & BS-64\\
Draco172015.4+575328.0 &  0.59960	   & 20.085 & 19.615 &  20.380  & 0.471 & ab    & BS-171\\
Draco172010.7+574559.1 &  0.60202$^{\dag}$ & 20.258 & 19.721 &  20.813  & 0.459 & ab    & BS-88\\
Draco172102.7+575251.0 &  0.60231	   & 20.146 & 19.587 &  20.444  & 0.834 & ab    & BS-80\\
Draco171922.9+574957.9 &  0.60241$^{\dag}$ & 20.160 & 19.308 &  20.785  & 0.445 & ab    & BS-137\\
Draco172116.6+575332.4 &  0.60257	   & 20.120 & 19.712 &  20.434  & 0.570 & ab,Bl & BS-26\\
Draco172057.2+575821.5 &  0.60336	   & 20.215 & 19.651 &  20.633  & 0.780 & ab    & BS-62\\
Draco172006.1+580206.8 &  0.60417	   & 20.143 &\nodata &  20.018  & 0.512 & ab    & BS-58\\
Draco172056.6+575352.9 &  0.60435	   & 20.232 & 19.656 &  20.652  & 0.381 & ab,Bl & BS-75\\
Draco172028.3+575701.2 &  0.60638	   & 20.171 & 19.679 &  20.637  & 0.688 & ab    & BS-103\\
Draco172009.2+574538.3 &  0.60730$^{\dag}$ & 20.229 & 19.599 &  20.575  & 0.937 & ab    & BS-89\\
Draco171922.9+575411.8 &  0.60813	   & 20.305 & 19.497 &  20.531  & 0.702 & ab    & \nodata\\
Draco171847.5+575305.3 &  0.60939	   & 20.156 & 19.497 &  20.597  & 0.715 & ab    & BS-60\\
Draco172051.2+575148.3 &  0.60998	   & 20.075 & 19.420 &  20.575  & 0.497 & ab    & BS-133\\
Draco172046.5+575742.0 &  0.61062	   & 20.192 & 19.606 &  20.586  & 0.569 & ab    & BS-63\\
Draco171951.3+574843.6 &  0.61159	   & 20.211 & 19.703 &  20.721  & 0.701 & ab    & BS-85\\
Draco172108.7+574746.7 &  0.61525$^{\dag}$ & 20.308 & 19.670 &  20.927  & 0.798 & ab    & BS-87\\
Draco172036.9+575213.1 &  0.61638	   & 20.136 & 19.442 &  20.469  & 0.500 & ab    & BS-40\\
Draco172009.3+575438.8 &  0.61688	   & 20.024 & 19.458 &  20.433  & 0.474 & ab,Bl & BS-68\\
Draco172055.1+574335.2 &  0.61701	   & 20.306 & 19.675 &  20.850  & 0.915 & ab    & \nodata\\
Draco172107.0+575942.5 &  0.61720	   & 20.077 & 19.383 &  20.324  & 0.740 & ab    & BS-95\\
Draco171835.1+575654.9 &  0.61790	   & 20.134 & 19.580 &  20.524  & 0.629 & ab    & BS-23\\
Draco171858.7+575257.0 &  0.61837	   & 20.044 & 19.456 & \nodata  & 0.878 & ab    & BS-14\\
Draco171945.6+575241.0 &  0.61864	   & 20.127 & 19.339 &  20.584  & 0.520 & ab    & \nodata\\
Draco172009.5+575957.7 &  0.61892	   & 20.293 & 19.814 &  20.668  & 0.537 & ab    & BS-7\\
Draco171917.6+575332.7 &  0.61954	   & 20.146 & 19.591 &  20.546  & 0.668 & ab    & BS-101\\
Draco171838.4+575238.4 &  0.61995	   & 20.196 & 19.472 &  20.579  & 0.776 & ab    & BS-20\\
Draco172008.5+574728.6 &  0.62023$^{\dag}$ & 20.003 & 19.510 &  20.533  & 0.690 & ab    & BS-49\\
Draco171851.9+574728.1 &  0.62029	   & 20.236 & 19.602 &  20.728  & 0.634 & ab    & \nodata\\
Draco172026.7+575427.7 &  0.62046	   & 20.215 & 19.737 &  20.624  & 0.632 & ab    & BS-106\\
Draco172053.1+575304.1 &  0.62066	   & 20.136 & 19.531 &  20.548  & 0.607 & ab    & BS-151\\
Draco172015.7+575417.4 &  0.62150	   & 20.767 & 19.622 &  20.703  & 0.859 & ab    & BS-71\\
Draco172040.6+575452.7 &  0.62158	   & 20.206 & 19.703 &  20.707  & 0.607 & ab    & BS-161\\
Draco172038.0+575531.2 &  0.62171	   & 20.176 & 19.493 &  20.590  & 0.541 & ab    & BS-162\\
Draco172107.1+575409.2 &  0.62505	   & 20.183 & 19.396 &  20.428  & 0.952 & ab    & BS-70\\
Draco172017.7+575200.9 &  0.62547	   & 19.997 & 19.349 &  20.396  & 0.770 & ab    & BS-36\\
Draco171845.3+575222.5 &  0.62589	   & 20.111 & 19.473 &  20.566  & 0.543 & ab    & BS-28\\
Draco172045.1+575128.0 &  0.62606	   & 20.201 & 19.522 &  20.591  & 0.499 & ab    & BS-164\\
Draco172029.9+580057.8 &  0.62629	   & 20.258 & 19.571 &  20.385  & 0.672 & ab    & BS-4\\
Draco171946.3+574744.4 &  0.62898$^{\dag}$ & 20.229 & 19.774 &  20.704  & 0.566 & ab    & BS-86\\
Draco172007.1+575949.7 &  0.62923	   & 20.205 & 19.503 &  20.468  & 0.599 & ab    & BS-98\\
Draco172053.3+575314.6 &  0.62974	   & 20.235 & 20.611 &  20.854  & 0.525 & ab    & BS-30\\
Draco171924.5+575336.3 &  0.63182	   & 20.200 & 19.639 &  20.617  & 0.562 & ab    & BS-19\\
Draco172024.2+575141.4 &  0.63182	   & 20.400 & 19.853 &  20.643  & 0.614 & ab    & BS-132\\
Draco172050.9+574517.2 &  0.63203$^{\dag}$ & 20.174 & 19.640 &  20.655  & 0.797 & ab    & BS-154\\
Draco171928.9+574916.6 &  0.63237$^{\dag}$ & 20.223 & 19.676 &  20.742  & 0.537 & ab    & BS-47\\
Draco172114.1+575435.9 &  0.63392	   & 20.193 & 19.660 &  20.586  & 0.524 & ab    & BS-128\\
Draco171942.6+575329.9 &  0.63957	   & 20.225 & 19.652 &  20.597  & 0.512 & ab    & BS-77\\
Draco172014.4+574402.3 &  0.64279	   & 20.111 & 19.539 &  20.721  & 0.559 & ab    & \nodata\\
Draco172008.9+575529.7 &  0.64325	   & 20.259 & 19.605 &  20.671  & 0.499 & ab,Bl & BS-160\\
Draco171955.4+574900.5 &  0.64495	   & 19.809 & 19.145 &  20.332  & 0.688 & ab    & BS-114\\
Draco172014.9+580146.6 &  0.64701	   & 20.313 & 19.036 &  20.569  & 0.720 & ab    & BS-3\\
Draco171858.8+575805.6 &  0.64741$^{\dag}$ & 20.208 & 19.553 & \nodata  & 0.436 & ab    & BS-66\\
Draco171905.6+575538.9 &  0.65139	   & 20.048 & 19.440 &  20.444  & 0.546 & ab    & \nodata\\
Draco171945.0+575418.1 &  0.65290$^{\dag}$ & 20.142 & 19.514 &  20.638  & 0.655 & ab    & BS-159\\
Draco172031.2+575737.5 &  0.65294	   & 20.178 & 19.612 &  20.532  & 0.654 & ab    & BS-158\\
Draco172029.3+575808.3 &  0.65345	   & 20.232 & 19.646 &  20.643  & 0.200 & ab    & \nodata\\
Draco172017.1+574641.0 &  0.65431$^{\dag}$ & 20.293 & 19.685 &  20.798  & 0.736 & ab    & BS-140\\
Draco171929.3+574159.4 &  0.66096	   & 20.164 & 19.586 &  20.541  & 0.471 & ab    & \nodata\\
Draco172013.2+575526.4 &  0.66261	   & 20.183 & 19.580 &  20.516  & 0.330 & ab    & BS-192\\
Draco172036.8+574820.6 &  0.66284	   & 20.107 & 19.442 &  20.551  & 1.146 & ab    & BS-172\\
Draco171931.8+575927.0 &  0.66353	   & 20.117 & 19.420 &  20.568  & 0.666 & ab    & \nodata\\
Draco171951.3+575321.1 &  0.66449	   & 20.193 & 19.536 &  20.636  & 0.686 & ab    & BS-127\\
Draco172025.9+580002.3 &  0.66465	   & 20.157 & 19.530 &  20.582  & 0.363 & ab    & BS-119\\
Draco172007.7+580140.5 &  0.66755	   & 20.205 & 19.607 &  20.591  & 0.610 & ab    & BS-167\\
Draco171948.8+575657.0 &  0.67368	   & 20.096 & 19.499 &  20.449  & 0.505 & ab    & BS-188\\
Draco171928.2+580043.3 &  0.67538	   & 20.230 & 19.672 &  20.667  & 0.360 & ab    & BS-149\\
Draco171905.7+575520.1 &  0.67602	   & 20.073 & 19.475 &  20.400  & 0.550 & ab    & BS-174\\
Draco172031.4+575303.0 &  0.67633$^{\dag}$ & 20.095 & 19.471 &  20.497  & 0.402 & ab,Bl & BS-150\\
Draco172021.6+575431.2 &  0.67650	   & \nodata& 19.625 &  20.624  &\nodata& ab    & BS-193\\
Draco172046.5+574818.9 &  0.67655	   & 20.163 & 19.500 &  20.817  & 0.347 & ab    & \nodata\\
Draco172051.8+575636.0 &  0.67955	   & 20.049 & 19.358 &  20.412  & 0.337 & ab    & BS-198\\
Draco172053.5+575704.9 &  0.68168	   & 20.157 & 19.596 &  20.547  & 0.585 & ab    & BS-125\\
Draco171944.0+575509.7 &  0.68413	   & \nodata& 19.442 & \nodata  &\nodata& ab    & BS-9\\
Draco171926.1+574851.4 &  0.68930$^{\dag}$ & 20.174 &\nodata &  20.703  & 0.471 & ab    & BS-187\\
Draco172018.9+580038.0 &  0.69487	   & 20.086 & 19.549 &  20.469  & 0.774 & ab    & BS-6\\
Draco172021.1+575219.1 &  0.73201	   & 20.024 & 19.459 &  20.497  & 0.319 & ab    & BS-81\\
Draco171944.8+575737.2 &  0.74359	   & 20.021 & 19.388 &  20.442  & 0.291 & ab    & BS-100\\
Draco172006.0+575349.4 &  0.76408	   & 19.966 & 19.473 &  20.392  & 0.473 & ab    & BS-27\\
Draco172039.0+575732.6 &  0.79062	   & 20.006 & 19.381 &  20.403  & 0.188 & ab    & \nodata\\
\enddata
\label{rrlyr}
\tablenotetext{a}{Names given in \citet{Baa61}.}
\tablenotetext{*}{Notes different period from that found by \citet{Baa61}.}
\tablenotetext{\dag}{Notes a star with no period in \citet{Baa61}.}
\end{deluxetable}

\begin{deluxetable}{crccccl}
\tabletypesize{\footnotesize}
\tablewidth{0pc}
\tablecaption{\sc Other Variables in Draco Dwarf Galaxy}
\tablehead{
\colhead{} & \colhead{$P$} & \colhead{$\langle V \rangle$} &\colhead{$\langle I \rangle$} & \colhead{$\langle B \rangle$} & \colhead{$\rm Amp_{V}$} & \colhead{} \\
\colhead{Name} & \colhead{($days$)} & \colhead{(mag)} & \colhead{(mag)} & \colhead{(mag)} & 
\colhead{(mag)} & \colhead{Comments$^{\rm a}$}}
\startdata
Draco172017.3+574817.3 &  0.068,0.073,0.079& 21.985& 21.468&  22.033  & 0.873 & SX Phe \\
Draco171906.2+574120.9 &  0.2435&   19.645 & 18.058&  20.751  & 0.266 & Ecl. Binary\\
Draco171906.4+574948.2 &  0.59229*& 18.846 & 18.368&  19.267  & 0.877 & Ceph, BS-134\\
Draco172017.8+575708.1 &  0.90085&  19.204 & 18.573&  19.640  & 0.679 & Ceph, BS-141\\
Draco171934.5+575849.6 &  0.93644&  19.043 & 18.439&  19.597  & 0.888 & Ceph, BS-157\\
Draco171918.4+575245.7 &  1.58993&  18.123 & 17.501&  18.542  & 0.460 & Ceph, BS-194\\
Draco171855.0+574733.1 &  5.1150&   17.707 & 16.586&  18.655  & 0.101 \\
Draco172037.4+575913.0 & \nodata&   17.051 & 15.584&  18.315  & \nodata \\
Draco172032.9+575144.2 & \nodata&   16.977 & 15.418&  18.430  & \nodata \\
Draco172040.2+575733.0 & \nodata&   16.488 & 14.927&  17.893  & \nodata \\
Draco172106.3+575410.9 & \nodata&   16.658 & 19.396&  20.428  & \nodata \\
Draco171929.5+575825.6 & \nodata&   16.855 & 15.279&  18.173  & \nodata \\
Draco171922.5+575404.2 & \nodata&   17.326 & 15.921&  18.932  & \nodata \\
Draco171935.5+575846.4 & \nodata&   17.161 & 15.622&  18.509  & \nodata \\
Draco171945.8+575626.1 & \nodata&   19.548 & 18.832&  19.783  & \nodata & BS-203\\
Draco171952.2+575909.3 & \nodata&   17.487 & 16.078&  18.690  & \nodata \\
Draco172120.1+580118.7 & \nodata&   16.102 & 15.239&  16.866  & \nodata \\
\enddata
\label{var}
\tablenotetext{a}{Type of variable and name given in \citet{Baa61}.}
\tablenotetext{*}{Notes different period from that found by \citet{Baa61}.}
\end{deluxetable}

\end{document}